\documentclass[aps,prd,floats,preprintnumbers,showpacs,nofootinbib]{revtex4}
\usepackage{epsfig}
\usepackage{amsmath}
\usepackage{amssymb}
\usepackage{amsthm}

\begin{document}

\preprint{NUHEP-TH/11-22}

\title{Constraining the (Low-Energy) Type-I Seesaw}

\author{Andr\'e de Gouv\^ea}
\affiliation{Northwestern University, Department of Physics \& Astronomy, 2145 Sheridan Road, Evanston, IL~60208, USA}

\author{Wei-Chih Huang}
\affiliation{Northwestern University, Department of Physics \& Astronomy, 2145 Sheridan Road, Evanston, IL~60208, USA}


\begin{abstract}

The type-I seesaw Lagrangian yields a non-generic set of active--sterile oscillation parameters --- the neutrino mass eigenvalues and the physical elements of the full mixing matrix are entwined. For this reason one is able to, in principle, test the model by performing enough measurements which are sensitive to neutrino masses and lepton mixing. We point out that for light enough right-handed neutrino masses --- less than 10~eV --- next-generation short-baseline neutrino oscillation experiments may be able to unambiguously rule out (or ``rule in'') the low energy seesaw as the Lagrangian that describes neutrino masses. These types of searches are already under consideration in order to address the many anomalies from accelerator neutrino experiments (LSND, MiniBooNe), reactor neutrino experiments (the ``reactor anomaly'') and others. In order to test the low-energy seesaw, it is crucial to explore different oscillation channels, including $\nu_e$ and $\nu_{\mu}$ disappearance and $\nu_{\mu}\leftrightarrow\nu_{\tau}$ appearance.

\end{abstract}

\maketitle

\section{Introduction}

The simplest way of accommodating nonzero neutrino masses is to add to the standard model particle content at least two Weyl fermions $N_i$ ($i=1,2,\ldots$)  which are not charged under the known $SU(3)_c\times SU(2)_L\times U(1)_Y$ gauge group. The most general renormalizable Lagrangian is very well-known and easy to write down. Gauge singlet fermions --- henceforth referred to as right-handed or sterile neutrinos --- are only allowed to couple to the standard model via a Yukawa interaction with the Higgs and lepton $SU(2)_L$ doublets. On the other hand, right-handed neutrinos are allowed to have nonzero Majorana masses. In summary, the most general renormalizable Lagrangian is
\begin{equation}
{\cal L}_{\nu} = {\cal L}_{\rm old}+{\cal L}_{\rm kin} - \frac{M_i}{2} N_iN_i - y^{\alpha i} L_{\alpha} N_i H  + H.c.,
\label{eq:lnu}
\end{equation}
where ${\cal L}_{\rm old}$ is the standard model Lagrangian in the absence of gauge singlet fermions, ${\cal L}_{\rm kin}$ are the kinetic energy terms for the $N_i$ fields, $\alpha=e,\mu,\tau$ are the flavor indices of the three lepton doublet fermion fields $L$, and $H$ is the Higgs boson doublet field. $y$ are the neutrino Yukawa couplings, while $M$ are the right-handed neutrino Majorana mass parameters. Note that we picked a weak basis for the $N$ states such that their Majorana mass matrix is diagonal.  

After electroweak symmetry breaking, Eq.~(\ref{eq:lnu}) describes, in general, $3+n$ ($n$ is the number of right-handed neutrinos) Majorana fermions --- neutrinos --- most of which are massive if $n\ge 2$. It is very easy to accommodate the universally accepted neutrino oscillation data by judiciously choosing values for $y$ and $M$. Assuming all $M$ to be of the same order of magnitude, current data rule out $1~{\rm neV}\lesssim M\lesssim 1$~eV \cite{deGouvea:2005er,deGouvea:2006gz,deGouvea:2009fp,Donini:2011jh}. All other values are allowed!

More information is required in order to test whether Eq.~(\ref{eq:lnu}) is the correct way of understanding neutrino masses. Theoretical considerations allow one to rule out $M\gtrsim 10^{15}$~GeV \cite{Maltoni:2000iq}, while a simple interpretation of the gauge hierarchy problem leads one to favor $M\lesssim 10^{7}$~GeV, assuming there are no other new states at or above the electroweak symmetry breaking scale \cite{Casas:2004gh}. Naturalness is not a good guide when it comes to picking a value for $M$ --- all values for $M$ are technically natural in the sense that in the limit where all $M$ vanish the global symmetries of Eq.~(\ref{eq:lnu}) are augmented by $U(1)_{L}$, global lepton number (see, for example, \cite{deGouvea:2005er}).\footnote{Lepton number, along with baryon number, is anomalous. Nonetheless, if $n=3$, $U(1)_{B-L}$ is a proper non-anomalous global symmetry of Eq.~(\ref{eq:lnu}) in the limit $M\to 0$.}   

Direct experimental probes of Eq.~(\ref{eq:lnu}) are possible for $M$ values below the TeV scale. Here we are mostly interested in $M$ values in the few to several eV range. In this mass-range, right-handed neutrinos can be ``seen'' in neutrino oscillation data. Most interesting, in this mass-range, there is the possibility of unambiguously testing Eq.~(\ref{eq:lnu}). The reason for this is that Eq.~(\ref{eq:lnu}) implies relations among some of the mixing angles that can be measured in neutrino oscillation experiments, as we discuss in detail in Sec.~\ref{sec:formalism}. In particular, given what is known about neutrino masses and lepton mixing today, simple lower bounds on different combinations of mixing parameters exist and, if these were to be violated by future experimental data, Eq.~(\ref{eq:lnu}) would be ruled out for certain values of $M$. We discuss some of these quantitative bounds in Sec.~\ref{sec:results}.

Outside of $M$ around a few to several eV, there are several interesting direct and indirect experimental bounds on Eq.~(\ref{eq:lnu}) which have been discussed in the literature. We briefly comment on these, and summarize our results in Sec.~\ref{sec:conclusions}.

\section{Formalism and Relations Among Elements of the Mixing Matrix}
\label{sec:formalism}

After electroweak symmetry breaking, the $(3+n)\times (3+n)$ neutrino Majorana mass matrix from Eq.~(\ref{eq:lnu}) is
\begin{equation}
m_{\nu}=\left(\begin{array}{cc} 0_{3\times 3} & m_D \\ m_D^T & M \end{array}
\right),
\label{eq:mnu}
\end{equation}
where $0_{3\times 3}$ stands for a $3\times 3$ zero matrix, $M$ is the $n\times n$ matrix of right-handed neutrino Majorana mass parameters and $m_D=y v$ is the $3\times n$ matrix of Dirac mass parameters. $v$ is the vacuum expectation value of the neutral component of $H$. Eq.~(\ref{eq:mnu}) is expressed in the so-called flavor basis: $\nu_e, \nu_{\mu}, \nu_{\tau}, \nu_{s_1}, \ldots,\nu_{s_n}$. This is related to the neutrino mass basis --- $\nu_1,\nu_2,\ldots,\nu_{3+n}$ --- via the $(3+n)\times (3+n)$ neutrino unitary mixing matrix $U$: 
$$
\nu_{\alpha}=U_{\alpha i}\nu_i,~~~~UU^{\dagger}=U^{\dagger}U=1,
$$
$\alpha=e,\mu,\tau,s_1,\ldots,s_n$, $i=1,2,\ldots,n+3$ such that
\begin{equation}
m_{\nu}=\left(\begin{array}{cc} 0_{3\times 3} & m_D \\ m_D^T & M \end{array}
\right)=U^*\left(\begin{array}{cccc} m_1 &  & &  \\  & m_2 & &  \\  &  & \ddots & \\ & & & m_{3+n}\end{array}\right)U^{\dagger},
\label{eq:flavor_mass}
\end{equation}
where $m_1,m_2,\ldots,m_{3+n}$ are the neutrino masses. It is instructive to express $U$ in terms of 4 submatrices,
\begin{equation}
U=\left(\begin{array}{cc} V_{\rm MNS} & \Theta \\ \Theta'^T & V_s \end{array}
\right),
\end{equation}
where $V_{\rm MNS}$ is a $3\times 3$ matrix that will be referred to as the ``active'' mixing matrix, $V_s$ is an $n\times n$ matrix that will be referred to as the ``sterile'' mixing matrix, $\Theta,\Theta'$ are $3\times n$ matrices and we will refer to $\Theta$ as the ``active--sterile'' mixing matrix. In the absence of sterile neutrinos,\footnote{In this case, of course, the neutrino masses would have to be generated through other means, or the neutrinos are Dirac fermions (all $M=0$).} $V_{\rm MNS}$ is the standard neutrino mixing matrix whose elements have currently been measured with different degrees of precision \cite{Nakamura:2010zzi}. In practice, unless otherwise noted, we will assume that the elements of $\Theta$ are parametrically smaller than those of $V_{\rm MNS}$ and that $V_{\rm MNS}$ is approximately unitary: $V_{\rm MNS}V_{\rm MNS}^{\dagger}\sim 1_{3\times 3}$.

In the absence of interactions beyond those in Eq.~(\ref{eq:lnu}), physics is invariant under rotations among the sterile states,
\begin{equation}
U\to \left(\begin{array}{cc} 1_{3\times 3} & 0_{3\times n} \\ 0_{n\times 3} & U_{n\times n} \end{array}
\right) U,
\label{eq:ssym}
\end{equation}
where $1_{3\times 3}$ is the $3\times 3$ identity matrix and $U_{n\times n}$ is any $n\times n$ unitary matrix. This invariance allows one to significantly reduce the number of physical parameters in $U$ and reveals that all physical phenomena are specified once the elements of $V_{\rm MNS}$ and $\Theta$ are known (these are invariant under Eq.~(\ref{eq:ssym})). In summary, the ``physical'' parameters of $U$ are $U_{ai}$, where $a=e,\mu,\tau$, $i=1,2,3,\ldots, 3+n$ the active neutrino flavors. $U_{si}$, $s=s_1,\ldots,s_n$ cannot be directly measured. For more details on how to parameterize $U$ in the case at hand see, for example, \cite{deGouvea:2008nm,deGouvea:2010iv,Asaka:2011pb,Blennow:2011vn}.

Eq.~(\ref{eq:flavor_mass}) further provides nontrivial relations among the elements of $U$ and the neutrino mass eigenvalues. In detail,
\begin{equation}
(m_{\nu})_{\alpha\beta}=\sum_i U^*_{\alpha i}U^*_{\beta i} m_i,
\label{eq:mab,U}
\end{equation}
where $\alpha,\beta=e,\mu,\tau,s_1,\ldots,s_n$. We now proceed as follows: we will choose as {\sl inputs} $m_i$ and the elements of $V_{\rm MNS}$ and try to predict the elements of $\Theta$. All the extra information comes from the vanishing ``active--active'' elements of $m_{\nu}$ --- $0_{3\times 3}$:
\begin{eqnarray}
&\sum_i U_{a i}U_{b i} m_i=0, \nonumber \\
&\sum_{i=1}^3 U_{a i}U_{b i} m_i + \sum_{i=1}^n U_{a, i+3}U_{b, i+3} m_{i+3}=0, \nonumber \\
&\sum_{i=1}^n \Theta_{a i}\Theta_{b i} m_{i+3} =-\sum_{i=1}^3 (V_{\rm MNS})_{ai}(V_{\rm MNS})_{bi} m_i, \label{eq:0const}
\end{eqnarray}
for all $a,b=e,\mu,\tau$. Note that we choose all $m_i$ to be real in order to re-express Eq.~({\ref{eq:mab,U}}) in terms of the elements of $U$, as opposed to $U^*$.

A few simplified scenarios illustrate the spirit of the results we will discuss below. For example, if there was only one active $\nu_a$ and one right-handed neutrino $\nu_s$, Eq.~(\ref{eq:0const}) would read
\begin{equation}
U_{a4}^2m_4 = -U_{a1}^2m_1, 
\end{equation}
where, in order to facilitate future comparison with more complicated scenarios, we label our masses $m_1$ and $m_4$. In this case $U$ can be chosen real and, if we further redefine the neutrino mass eigenstates in such a way that $m_1\to-m_1$, it is easy to see that the only parameter\footnote{We choose the standard parameterization for the $2\times 2$ mixing matrix, $U_{a1}=U_{s4}=\cos\theta$, $U_{a2}=-U_{s1}=\sin\theta$.} necessary to parameterize $U$ is uniquely determined by the neutrino mass eigenvalues: $\tan^2\theta=m_1/m_4$. This scenario could be tested if one were to independently measure the two neutrino masses and the one leptonic mixing angle.  

In the case of one active and two sterile states, Eq.~(\ref{eq:0const}) reads (after redefining $m_1\to-m_1$)
\begin{equation}
U_{a4}^2m_4+U_{a5}^2m_5 = U_{a1}^2m_1.
\end{equation}
In this case, it is possible to obtain {\sl lower bounds} for the active--sterile mixing angles as a function of $U_{a1}$ and the masses. A formal solution to the constraint above is
\begin{eqnarray}
U_{a4}^2=U_{a1}^{2}\frac{m_1}{m_4}\cos^2\zeta, \\
U_{a5}^2=U_{a1}^{2}\frac{m_1}{m_5}\sin^2\zeta.
\end{eqnarray}
where $\zeta\in\mathbb{C}$. One can choose $U_{a1}$ real and easily show 
\begin{equation}
|U_{a4}|^{2}m_4+|U_{a5}|^{2}m_5 \ge |U_{a1}|^{2}m_1.
\end{equation}
The inequality is saturated for $\Im(\zeta)=0$. If either entry of the active--sterile part of the mixing matrix vanishes, the magnitude of the other one is guaranteed to be equal to $U_{a1}\sqrt{\frac{m_1}{m_{4,5}}}$. The careful reader will note that unitarity imposes other constraints, including, $|U_{a1}|^2+|U_{a4}|^2+|U_{a5}|^2=1$. As mentioned in passing earlier, we are working under the assumption that $m_{4,5,\ldots}\gg m_{1,2,3}$ and the elements of $V_{\rm MNS}$ are parametrically larger than those of $\Theta$. In this case, this assumption translates into $U_{a1}=1+O(m_1/m_{m_{4,5}})^2$.

In the case of two active states $\nu_a$ and $\nu_b$ and one sterile state, Eq.~(\ref{eq:0const}) reads (after redefining $m_4\to-m_4$)
\begin{eqnarray}
U_{a4}^2m_4 = U_{a2}^2m_2, \\
U_{b4}^2m_4 = U_{b2}^2m_2, \\
U_{a4}U_{b4}m_4=U_{a2}U_{b2}m_2,
\end{eqnarray}
where we took into account that, in this case, one of the neutrino masses (here $m_1$) vanishes. The three equations above translate into 
\begin{eqnarray}
U_{a4}^2=U_{a2}^2\frac{m_2}{m_4}, \\
U_{b4}^2=U_{b2}^2\frac{m_2}{m_4}.
\end{eqnarray}
Here, all elements of $U$ can be chosen real. Requiring $m_4\gg m_2$ we can approximate $U_{a2}^2=1-U_{b2}^2=\sin^2\vartheta$, and it is easy to see that the active--sterile entries of the mixing matrix are well-define functions of the active mixing angle $\vartheta$ and the neutrino masses, $m_2,m_4$. By measuring ``active--active'' oscillations governed by the small $\Delta m^2_{12}=m_2^2$ mass-squared difference one can in principle measure $\vartheta$ and $m_2^2$ and hence express the oscillation probabilities at short distances as a function of only one new parameter ($m_4$). For example, for an ultrarelativistic beam of neutrinos with energy $E$ and short baselines $L$ ($m_2^2L/E\ll 1$),
\begin{eqnarray}
&P_{aa}\sim1-4\sin^2\vartheta\left(\frac{m_2}{m_4}\right)\sin^2\left(\frac{m_4^2L}{4E}\right), \\
&P_{bb}\sim1-4\cos^2\vartheta\left(\frac{m_2}{m_4}\right)\sin^2\left(\frac{m_4^2L}{4E}\right), \\
&P_{ab}\sim \sin^22\vartheta\left(\frac{m_2}{m_4}\right)^2\sin^2\left(\frac{m_4^2L}{4E}\right).
\end{eqnarray} 

For more complicated cases, including the physically relevant case of three active neutrino flavors, it is convenient to formally solve Eq.~(\ref{eq:0const}), which can first be re-expressed as a matrix equation. After redefining the sign of the light neutrino masses ($m_1\to-m_1$, etc), Eq.~(\ref{eq:0const}) reads
\begin{equation}
(\Theta\sqrt{m_{\rm heavy}})(\Theta\sqrt{m_{\rm heavy}})^T=(V_{\rm MNS}\sqrt{m_{\rm light}})(V_{\rm MNS}\sqrt{m_{\rm light}})^T,
\label{eq:matrix}
\end{equation}
where $m_{\rm heavy}={\rm diag}(m_4,\ldots,m_{3+n})$,  $m_{\rm light}={\rm diag}(m_1,m_2,m_3)$. The equation above allows one to write
\begin{equation}
\Theta\sqrt{m_{\rm heavy}}=V_{\rm MNS}\sqrt{m_{\rm light}}R,
\label{eq:casas_ibarra}
\end{equation}
where $R$ is a complex matrix. This very useful solution was first presented in \cite{Casas:2001sr}. In the case $n=3$, $R$ is a complex orthogonal matrix, $RR^T=R^TR=1$. Here, we will also be interested in the case $n=2$ when $R$ ``contains'' a complex orthogonal matrix,\footnote{In the ``1+2'' case discussed above, $R_{1\times 2}=(\cos\zeta~~~\sin\zeta)$ and $RR^T=1$, while in the ``2+1'' case $R_{2\times 1}=(0~~~1)^T$. In the latter case, $RR^T={\rm diag}(0,1)$.}
\begin{equation}
R_{3\times 2}=\left(\begin{array}{cc} 0 & 0 \\ \cos\zeta & \sin\zeta \\ -\sin\zeta & \cos\zeta \\ \end{array}\right)~~~{\rm or}~~~\left(\begin{array}{cc} \cos\zeta & \sin\zeta \\ -\sin\zeta & \cos\zeta \\ 0 & 0 \end{array}\right).
\label{eq:R32}
\end{equation}
We choose the vanishing neutrino mass eigenvalue to be $m_1=0$ or $m_3=0$, respectively, and $\zeta$ is a complex number. We draw attention to the decisive role played by the zero eigenvalue in rendering Eq.~(\ref{eq:casas_ibarra}) a solution to Eq.~(\ref{eq:matrix}).

Before moving on to the physically relevant cases, we quickly discuss the case of two active states and two sterile states. Eq.~(\ref{eq:casas_ibarra}) reads
\begin{equation}
\left(\begin{array}{cc} U_{a4}\sqrt{m_4} & U_{a5}\sqrt{m_5} \\ U_{b4}\sqrt{m_4} & U_{b5}\sqrt{m_5} \end{array}\right)=
\left(\begin{array}{cc} \cos\vartheta\sqrt{m_1}e^{i\varphi}  & \sin\vartheta\sqrt{m_2}\\ -\sin\vartheta\sqrt{m_1}e^{i\varphi} & \cos\vartheta\sqrt{m_2} \end{array}\right)
\left(\begin{array}{cc} \cos\zeta & \sin\zeta \\ -\sin\zeta & \cos\zeta \\ \end{array}\right),
\label{eq:2+2}
\end{equation}
where $\zeta\in\mathbb{C}$ and we assumed $m_{4,5}\gg m_{1,2}$, and the active--active mixing matrix  is approximately unitary and parameterized by one mixing angle $\vartheta$ and one ``Majorana'' phase $\varphi$. Overall, the $4\times 4$ mixing matrix is parameterized in terms of four real parameters, $\vartheta,\varphi$, and the real and imaginary parts of $\zeta$.

In order to separate the impact of the relations among the different elements of $U$ on the possible values of active--sterile ``mixing angles,'' it is convenient to work with the dimensionless objects $X_{aj}\equiv U_{aj}\sqrt{m_j}/\sqrt{m_2}$ ($j=4,5$) and $c\equiv\sqrt{m_1}/\sqrt{m_2}e^{i\varphi}$, where $m_2$ is defined as the largest of $m_{1,2}$. Hence
\begin{equation}
\left(\begin{array}{cc} X_{a4} & X_{a5} \\ X_{b4} & X_{b5} \end{array}\right)=
\left(\begin{array}{cc} c\cos\vartheta\cos\zeta-\sin\vartheta\sin\zeta & c\cos\vartheta\sin\zeta+\sin\vartheta\cos\zeta \\ -c\sin\vartheta\cos\zeta- \cos\vartheta\sin\zeta & -c\sin\vartheta\sin\zeta+\cos\vartheta\cos\zeta \end{array}\right).
\end{equation}

In general, one can always pick $\zeta$ such that one of the $X_{aj}$ vanishes. When that happens, however, none of the other three vanish.\footnote{We are ignoring special values of, or special relations between, $\vartheta$ and $c$. These are, after all, independent, a priori uncorrelated, parameters.} For example, $X_{a4}=0$ implies $\tan\zeta=c/\tan\vartheta$. Under these circumstances, all other elements of $X_{aj}$ are uniquely determined functions of the three (real) active neutrino parameters, $\vartheta,c$.

\setcounter{equation}{0} \setcounter{footnote}{0}
\section{Quantitative Upper Bounds}
\label{sec:results}

Here we quantitatively discuss the expectations for the values of the elements of $\Theta$ in the physically interesting  case of three active neutrinos and $n=2$ or $n=3$. The $n=2$ case is both simpler to analyze and ``more minimal,'' so we devote most of our presentation to it. We first quickly summarize some of the inputs we will use throughout. 

\subsection{Active Oscillation Parameters}

Current experimental data constrain all the elements of $V_{\rm MNS}$ and two mass-squared differences,\footnote{We pick the ``standard'' parameterization for the active--active mixing matrix  and the mass-squared differences, along with the ``standard'' definition of the neutrino mass eigenvalues. For more details see \cite{Nakamura:2010zzi} and the many references therein.} $\Delta m^2_{12}$ and $|\Delta m^2_{13}|$. Throughout we assume $\sin^2\theta_{12}=0.3$, $\sin^2\theta_{23}=0.5$, $\Delta m^2_{12}=7.6\times 10^{-5}$~eV$^2$ and $|\Delta m^2_{13}|=2.4\times 10^{-3}$~eV$^2$, in agreement with recent analysis of all neutrino data \cite{Schwetz:2011zk,Fogli:2011qn}. Unless otherwise noted, we will assume $\sin^2\theta_{13}=0.01$. In summary, we assume
\begin{equation}
V_{\rm MNS}=\left(\begin{array}{ccc}
0.83 & 0.55 & 0.1e^{i\delta} \\ -0.39-0.06e^{-i\delta} & 0.59-0.04e^{-i\delta} & 0.70 \\ 0.39-0.06e^{-i\delta} & -0.59-0.04e^{-i\delta} & 0.70
\end{array}\right)\left(\begin{array}{ccc} e^{i\psi} & 0 & 0 \\ 0 & e^{i\phi} & 0 \\ 0 & 0 & 1 \end{array}\right).
\label{MNS_data}
\end{equation}
The CP-odd phase $\delta$ is currently only very poorly constrained and will be taken as a free parameter when relevant. Note that, since $\theta_{13}$ is known to be small, the impact of the uncertainty in $\delta$ on the magnitude of the elements of $V_{\rm MNS}$ is relatively small. The so-called Majorana phases $\psi,\phi$ will be taken as free parameters when relevant. When one of the mass eigenvalues vanishes, only one Majorana phase is physical.\footnote{If $m_1=0$, $\phi$ can be chosen as the physical Majorana phase. In the case $m_3=0$, one can choose $\phi-\psi$ as the physical Majorana phase.}

In the case of a so-called normal neutrino mass hierarchy, we express all neutrino masses as functions of $m_1$: $m_2^2=m_1^2+\Delta m^2_{12}$, $m_3^2=m_1^2+\Delta m^2_{13}$. Note that all masses are defined as real and positive. The square-root of the matrix $m_{\rm light}$ is
\begin{equation}
\sqrt{m_{\rm light}}=\sqrt{m_3}\left(\begin{array}{ccc} \sqrt{\frac{m_1}{m_3}} & 0 & 0 \\ 0 & \sqrt{\frac{m_2}{m_3}}  & 0 \\ 0 & 0 & 1 \end{array}\right)~~~\begin{array}{c} \longrightarrow \\  m_1\to 0 \\ \longrightarrow \end{array}~~~
\sqrt{m_{\rm light}}=
0.22\sqrt{\rm eV}\left(\begin{array}{ccc} 0 & 0 & 0 \\ 0 & 0.42  & 0 \\ 0 & 0 & 1 \end{array}\right).
\end{equation}

In the case of a so-called inverted neutrino mass hierarchy, we express all neutrino masses as functions of $m_3$: $m_1^2=m_3^2-\Delta m^2_{13}$ and $m_2^2=m_3^2-\Delta m^2_{13}+\Delta m^2_{12}$. We remind readers that $\Delta m^2_{13}$ is, in this case, negative. The square-root of the matrix $m_{\rm light}$ is
\begin{equation}
\sqrt{m_{\rm light}}=\sqrt{m_2}\left(\begin{array}{ccc} \sqrt{\frac{m_1}{m_2}} & 0 & 0 \\ 0 & 1  & 0 \\ 0 & 0 & \sqrt{\frac{m_3}{m_2}} \end{array}\right)~~~\begin{array}{c} \longrightarrow \\  m_3\to0 \\ \longrightarrow \end{array}~~~
\sqrt{m_{\rm light}}=
0.22\sqrt{\rm eV}\left(\begin{array}{ccc} 0.99 & 0 & 0 \\ 0 & 1  & 0 \\ 0 & 0 & 0 \end{array}\right).
\end{equation}

Similar to what was done in the end of the last section, we will be mostly interested in computing bounds that are independent from the values of the heavy masses. With that in mind, we define the matrix $X\equiv\Theta\sqrt{m_{\rm heavy}}/\sqrt{m_3}$ (normal hierarchy) or $X\equiv\Theta\sqrt{m_{\rm heavy}}/\sqrt{m_2}$ (inverted hierarchy). It is trivial to convert between a bound on $X_{ai}$ and $U_{ai}$. For example, for a normal hierarchy, $U_{e5}=X_{e5}\sqrt{m_3/m_5}$\footnote{it is convenient to label the elements of $X_{ai}$ such that $a=e,\mu,\tau$ and $j=4,5,\ldots 3+n$. We do that henceforth.} and, for an inverted hierarchy, $U_{\tau 4}=X_{\tau4}\sqrt{m_2/m_4}$.

With neutrino oscillation experiments one can study, in principle, the disappearance probability of active neutrinos, $P_{aa}$, or the conversion probability of one active flavor into another, $P_{ab}$ ($a,b=e,\mu,\tau$). These, in turn, are proportional to mass-squared differences and products of elements of the mixing matrix. Here, we will be interested in next-generation short baseline experiments. Short-baseline, here, means values of $L$ such that $|\Delta m^2_{13}|L/E\ll 1$, so ``atmospheric'' induced oscillations (and, hence, ``solar'' induced oscillations as well) can be safely ignored. Under these circumstances one is only sensitive to the large mass-squared differences. In the limit $m_{\rm heavy}\gg m_{\rm light}$, at short baselines, 
we will assume one is able to measure all different oscillation frequencies, proportional to $m_j^2$, $j=4,5,6$, and the following combinations of mixing elements: $|U_{aj}|^2$, $j=4,5,6$ (from disappearance) $U_{aj}U_{bj}^*$, $j=4,5,6$ (from appearance).\footnote{Needless to say, experimentally, this is a formidable task at best!} Henceforth, unless otherwise noted, we will concentrate on placing bounds on $|X_{aj}|^2$ and $|X_{aj}X_{bj}^*|$, for different $a,b=e,\mu,\tau$.

Before proceeding, we repeat our conventions and the approximations that go into our results. We assume there are three active and $n=2$ or 3 sterile neutrinos. After electroweak symmetry breaking, the neutrinos ``split'' into three light, mostly active neutrinos with masses $m_1,m_2,m_3$ and $n$ heavy, mostly sterile neutrinos with masses $m_4,m_5,\ldots,m_{3+n}$. We also assume that  $V_{\rm MNS}$ is approximately unitary and that its elements are known, except for the CP-odd phases (``Dirac'' and ``Majorana''). As far as the ``scale-free'' active--sterile mixing matrix $X$ is concerned, our approximations are safe if we restrict  all $X_{aj}\lesssim 1$. This way, all $\Theta_{aj}^2$ elements are constrained to be less than or of order $m_{\rm light}/m_{\rm heavy}$, which is small as long as we stick to $m_{\rm heavy}$ values above 1~eV and stay away from quasi-degenerate values for $m_1,m_2,m_3$. 

\subsection{$n=2$}

In the case of only two right-handed neutrinos, the lightest neutrino mass is zero and the $R$ matrix is given by Eq.~(\ref{eq:R32}). For a normal neutrino mass hierarchy,
\begin{equation}
X_{\rm normal} = \left(\begin{array}{cc}
0.23e^{i\phi} & 0.1e^{i\delta} \\ (0.25-0.02e^{-i\delta})e^{i\phi} & 0.70 \\ -(0.25+0.02e^{-i\delta})e^{i\phi} & 0.70
\end{array}\right)
\left(\begin{array}{cc} \cos\zeta & \sin\zeta \\ -\sin\zeta & \cos\zeta \\ \end{array}\right). \label{X_norm}
\end{equation}
On the other hand, for an inverted mass hierarchy,
\begin{equation}
X_{\rm inverted}=
\left(\begin{array}{cc}
0.83e^{i\psi} & 0.55 \\ -(0.39+0.06e^{-i\delta})e^{i\psi} & 0.59-0.04e^{-i\delta} \\ (0.39-0.06e^{-i\delta})e^{i\psi} & -0.59-0.04e^{-i\delta} 
\end{array}\right)
\left(\begin{array}{cc} \cos\zeta & \sin\zeta \\ -\sin\zeta & \cos\zeta \\ \end{array}\right). \label{X_inv}
\end{equation}

Before discussing bounds on the different elements of $X$, we discuss two interesting and potentially relevant examples. For $\zeta=3/4\pi+i$, $\delta=6/5\pi$, $\phi=\pi/2$ and a normal mass hierarchy,
\begin{equation}
X_{\rm normal} = \left(\begin{array}{cc}
0.41e^{-0.66i} & 0.45e^{1.03i} \\ 0.62e^{2.67i} & 0.61e^{-2.62i} \\ 1.27e^{2.44i} & 1.26e^{-2.41i}
\end{array}\right).
\label{eq:fit_normal}
\end{equation}
On the other hand, for $\zeta=2/3\pi+0.3i$, $\delta=0$, $\psi=\pi/2$, and an inverted mass hierarchy,
\begin{equation}
X_{\rm inverted} = \left(\begin{array}{cc}
0.44e^{-2.24i} & 0.62e^{1.83i} \\ 0.69e^{2.66i} & 0.66e^{-2.14i} \\ 0.71e^{-0.39i} & 0.60e^{0.89i}
\end{array}\right).
\label{eq:fit_inv}
\end{equation}
Both of these are consistent with the best ``3+2'' fits to the so-called short-baseline anomalies from LSND \cite{Aguilar:2001ty}, MiniBooNE \cite{AguilarArevalo:2007it,AguilarArevalo:2008rc,AguilarArevalo:2010wv} and reactor data \cite{Mention:2011rk}. From \cite{Kopp:2011qd}, for $m_4^2=0.5$~eV$^2$ and $m_5=0.9$~eV$^2$, the best fit point occurs when $|X_{e4}|=0.49$, $|X_{e5}|=0.62$, $|X_{\mu4}|=0.65$, $|X_{\mu5}|=0.66$ and ${\rm arg}(X_{e4}^*X_{\mu4}X_{e5}^*X_{\mu5})=5.0$. These are qualitatively in agreement with the matrices above.\footnote{The careful reader will note that the relevant relative phases in the examples are on the small side.} In the case of a normal mass hierarchy, a good fit can only be obtained if $\zeta$ has an order one imaginary part. For real $\zeta$, $|X_{e4,5}|$ are always too small. In the case of an inverted hierarchy, one can qualitatively ``fit'' the short-baseline anomalies even for real $\zeta$.  We note that a qualitatively similar analysis of the short-baseline anomalies was presented in \cite{Giunti:2011gz}.

Moving away from trying to explain the short-baseline anomalies, it is easy to see that, in general, for either mass hierarchy, at most one $X_{aj}$ can vanish.\footnote{Exceptions exist, of course, but these are entirely driven by the specific values of entries of $V_{\rm MNS}$. For example, for $U_{e3}=0$ and $U_{\tau3}=U_{\mu3}$ (vanishing $\theta_{13}$, maximal ``atmospheric'' mixing), it is easy to see that, for an inverted mass hierarchy, $|X_{\mu4,5}|=|X_{\tau 4,5}|$ such that if one picks a value of $\zeta$ that leads to a vanishing $X_{\mu4}$, $X_{\tau4}$ would also automatically vanish.} The situation here is similar to the $2+2$ case discussed in the previous section. This means that if some $|X_{aj}|$ are constrained to be very small, the scenarios under consideration will predict that other $|X_{aj}|$ are larger than some lower bound. It also implies that if one can constrain two different $|X_{aj}|$ to be smaller than some amount, the model can be conclusively ruled out.

\begin{figure}
\includegraphics[width=0.6\textwidth]{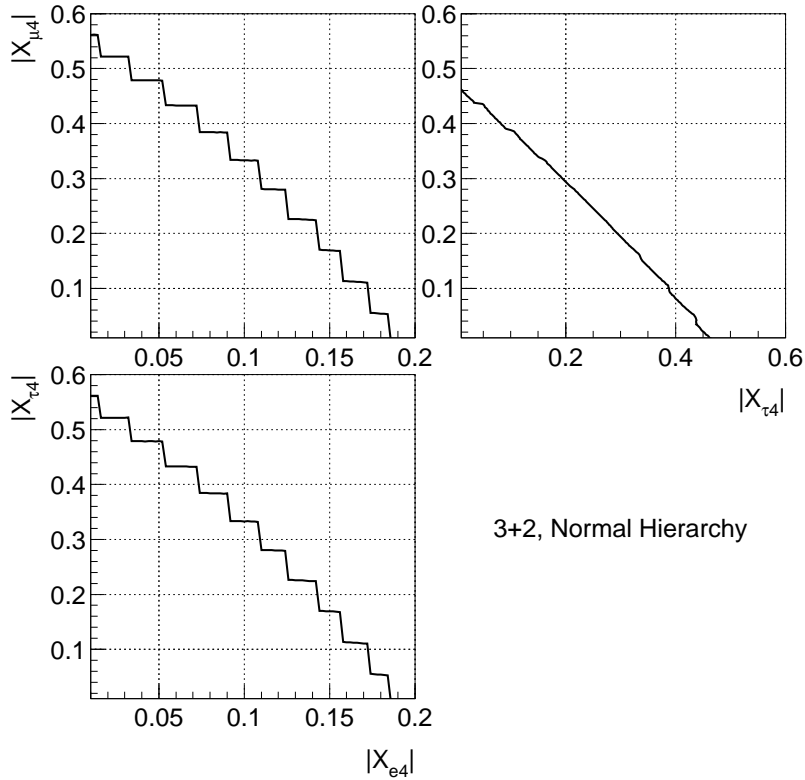}
\caption{Boundary of the lowest allowed values of mixing parameters in different $|X_{a4}|\times |X_{b4}|$ planes, $a,b=e,\mu,\tau$, in the case of a normal neutrino mass hierarchy and $n=2$. Different $X$ are defined in Eq.~(\ref{X_norm}). The  regions located to the left/bottom of the curves are not accessible.}
\label{X4s_normal}
\end{figure}

\begin{figure}
\includegraphics[width=0.6\textwidth]{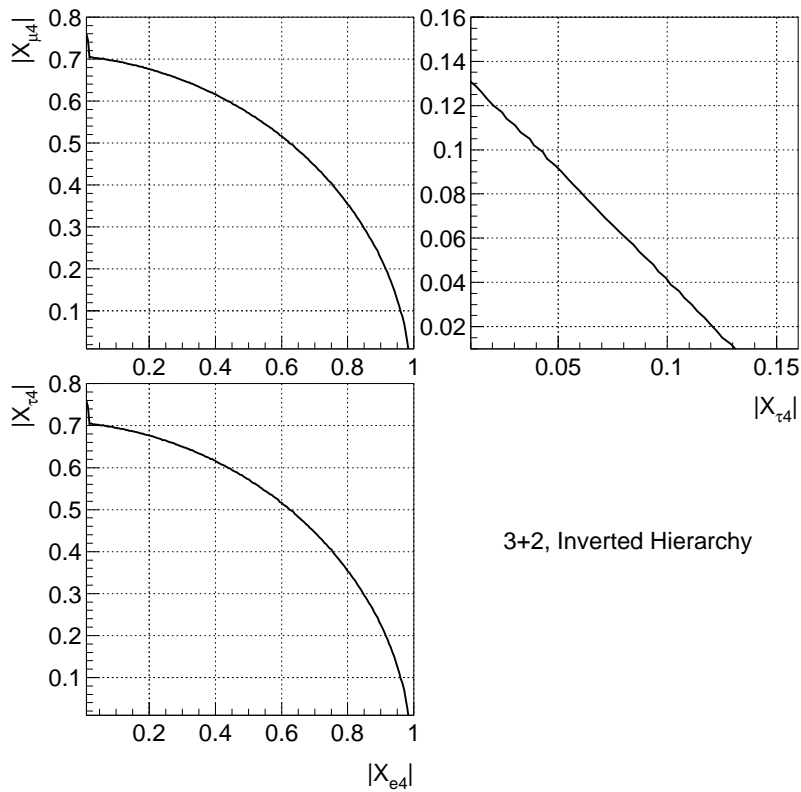}
\caption{Boundary of the lowest allowed values of mixing parameters in different $|X_{a4}|\times |X_{b4}|$ planes, $a,b=e,\mu,\tau$, in the case of an inverted neutrino mass hierarchy and $n=2$. Different $X$ are defined in Eq.~(\ref{X_inv}). The  regions located to the left/bottom of the curves are not accessible.}
\label{X4s_inverted}
\end{figure}

We begin by discussing the most conservative scenario: $m_4\ll m_5$. In this case, all practical information concerning testing the model will be provided by $X_{a4}$, $a=e,\mu,\tau$. When $m_4\ll m_5$, all $U_{a5}$ are parametrically much smaller than $U_{a4}$ (by a factor $\sqrt{m_4/m_5}$) and the only way to unambiguously test this model is to look for neutrino oscillations governed by the $m_4$-induced oscillation length. Figures~\ref{X4s_normal} and \ref{X4s_inverted} depict, for a normal and inverted hierarchy, respectively, the lowest accessible values of pairs of $|X_{e4}|$, $|X_{\mu4}|$ and $|X_{\tau4}|$ once one allows for all possible values\footnote{We numerically restrict the magnitude of the imaginary part of $\zeta$ during our exploration of the parameter space in order to satisfy the constraint $|X_{aj}|\lesssim 1$.} of $\zeta$, $\delta$, $\phi$ or $\psi$, respectively, as defined in Eqs.~(\ref{X_norm},\ref{X_inv}). The figures allow one to quickly confirm some of the statements made above. At most one of $|X_{a4}|$ vanishes for any point in the parameter space and there are ``forbidden regions'' (here we highlight the ones in the lower--left-hand side of the curves) in the space of the $|X_{a4}|$. If neutrino oscillation data were to limit these mixing parameters to within these forbidden regions, we would be able to rule out the model. It should be clear that by `model' we refer not only to a low-energy seesaw Lagrangian with two right-handed neutrinos, but also to the mass of the lightest of the mostly sterile states, in this case $m_4$. Next, we discuss two concrete examples (i.e., constraints for fixed values of $m_4$).

When $m_4=1$~eV, current electron neutrino disappearance data constrain $|X_{e4}|<0.45$ at the 90\% confidence level (C.L.) \cite{Declais:1994su}, while muon neutrino disappearance data constrain $X_{\mu4}<0.64$ at the 90\% C.L. \cite{Adamson:2011ku} (see also \cite{Mahn:2011ea}). Furthermore, searches for $\nu_{\mu}\to\nu_e$ reveal that $|X_{e4}||X_{\mu4}|<0.46$ at the 90\% C.L. \cite{AguilarArevalo:2007it}. Figs.~\ref{X4s_normal} and \ref{X4s_inverted} reveal that the $n=2$ scenarios are consistent with these results for either neutrino mass hierarchy. 

In the case of a normal neutrino mass hierarchy, the current data allow either $|X_{e4}|$ or $|X_{\mu4}|$ to be vanishingly small, in which case the other mixing element ($|X_{\mu4}|$ or $|X_{e4}|$) is close to current experimental constraints. In the case of an inverted neutrino mass hierarchy, the situation is much more constrained. The constraints spelled out above translate into, at the 90\% C.L., $|X_{e4}|>0.34$, $|X_{\mu4}|>0.60$, and $|X_{\tau4}|>0.61$. These, in turn, allow one to make very specific predictions concerning short-baseline oscillation experiments sensitive to a new mass-squared difference aroud 1~eV$^2$. These include the observation of \ldots
\begin{itemize}
\item $\nu_e$ disappearance with an associated effective mixing angle $\sin^22\vartheta_{ee}>0.02$. An interesting new proposal to closely expose the Daya Bay detectors to a strong $\beta$-emitting source would be sensitive to $\sin^22\vartheta_{ee}\gtrsim0.04$ \cite{Dwyer:2011xs};
\item $\nu_{\mu}$ disappearance with an associated effective mixing angle $\sin^22\vartheta_{\mu\mu}>0.07$, very close to the most recent MINOS lower bound;
\item $\nu_{\mu}\leftrightarrow\nu_e$ transitions with an associated effective mixing angle $\sin^2\vartheta_{e\mu}>0.0004$;
\item $\nu_{\mu}\leftrightarrow\nu_{\tau}$ transitions with an associated effective mixing angle $\sin^2\vartheta_{\mu\tau}>0.001$. A $\nu_{\mu}\to\nu_{\tau}$ appearance search sensitive to probabilities larger than 0.1\% for a mass-squared difference of 1~eV$^2$ would definitively rule out $m_4=1$~eV if the neutrino mass hierarchy is inverted.  
\end{itemize}

When $m_4=10$~eV,  current electron neutrino disappearance data constrain $|X_{e4}|<2.85$ at the 90\% C.L. \cite{Apollonio:2002gd}, while muon neutrino disappearance data constrain $|X_{\mu4}|<1.01$ at the 90\% C.L. \cite{Stockdale:1984cg}. Furthermore, searches for $\nu_{\mu}\to\nu_e$ reveal that $|X_{e4}||X_{\mu4}|<2.9$  \cite{Astier:2003gs} and those for $\nu_{\mu}\to\nu_{\tau}$ reveal that $|X_{\mu4}||X_{\tau4}|<1.8$, all at the 90\% C.L. \cite{Astier:2001yj}. Figs.~\ref{X4s_normal} and \ref{X4s_inverted} reveal that the $n=2$ scenarios are consistent with these results for either neutrino mass hierarchy. It is not surprising that the bounds on $X_{a4}$ are weaker than those for $m_4=1$~eV given that $|U_{a4}|^2\propto|X_{a4}|^2/m_4$. For $m_4=10$~eV and all $|X_{a4}|=0.5$, the related effective mixing angles are $\sin^22\vartheta_{aa}=0.005$ (appearance experiments) and $\sin^22\vartheta_{ab}=6\times 10^{-6}$. In order to unambiguously rule out $m_4=10$~eV and $n=2$, one needs to either probe $\nu_e$ or $\nu_{\mu}$ disappearance at the sub-percent level or look for appearance processes at the several parts-per-million level. Both appear very challenging, but the possibility of very sensitive $\nu_{\mu}\to\nu_{\tau}$ searches is currently under investigation (see, for example, \cite{Alonso:2010wu}). 

\begin{figure}
\includegraphics[width=0.6\textwidth]{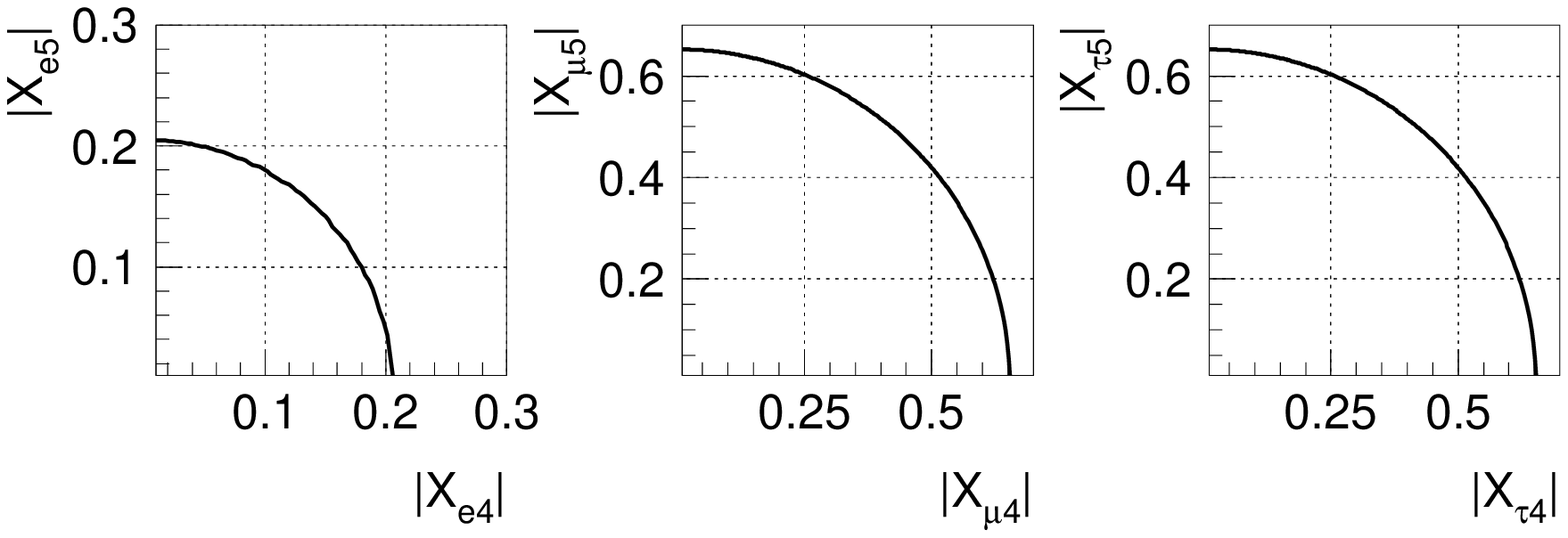}
\includegraphics[width=0.6\textwidth]{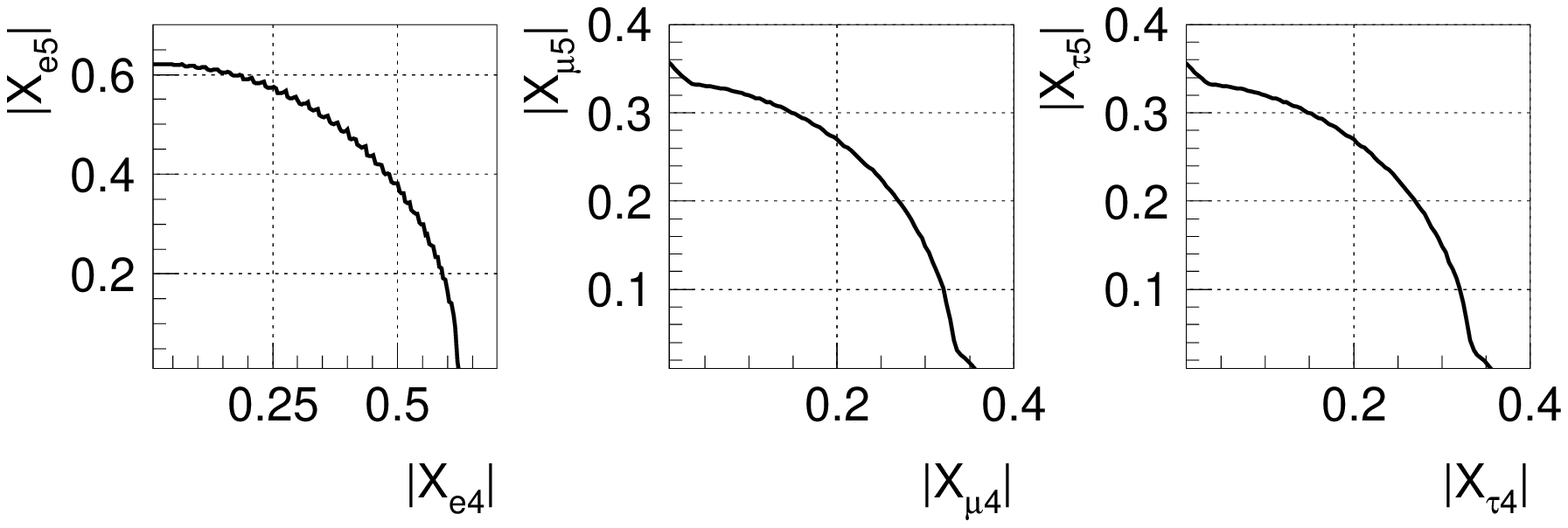}
\caption{Boundary of the lowest allowed values of mixing parameters in different $|X_{a4}|\times |X_{a5}|$ planes, $a=e,\mu,\tau$, in the case of a normal (top) or inverted (bottom) neutrino mass hierarchy, and $n=2$. Different $X$ are defined in Eqs.~(\ref{X_norm},\ref{X_inv}). The  regions located to the left/bottom of the curves are not accessible.}
\label{X4_X5}
\end{figure}

In the less conservative case where both $m_4$ and $m_5$ are ``small'' (say, below tens of eV) both new oscillation frequencies are accessible to searches for neutrino oscillations at short baselines and the $n=2$ low-energy seesaw scenario is much more constrained. Figure~\ref{X4_X5} depicts, for a normal and inverted hierarchy, the lowest accessible values of the magnitude of the active--sterile entries of the neutrino mixing matrix in the $|X_{a4}|\times |X_{a5}|$ plane, as defined in Eqs.~(\ref{X_norm},\ref{X_inv}). In our scan of the parameter space, we restricted the imaginary part of $\zeta$ to $\Im(\zeta)\in[-1,1]$. As advertised earlier, only at most one of the $X_{aj}$ can vanish for $j=4,5$ and $a=e,\mu,\tau$. 

The curves in Fig.~\ref{X4_X5} are easy to understand qualitatively. Eq.~(\ref{eq:0const}), for $a=b$ and $n=2$ implies
\begin{equation}
|X_{a4}|^2+|X_{a5}|^2\ge C_{a},
\label{eq:bound_n2}
\end{equation}
where $C_{a}$ are positive constants that depend only on active neutrino parameters. We will further explore this inequality in the next subsection. Hence one expects the boundaries in Fig.~\ref{X4_X5} to resemble quarter-circles. This is not the case for $a=\mu,\tau$ and an inverted neutrino mass hierarchy. The reason is that in order to saturate the bound in Eq.~(\ref{eq:bound_n2}), the magnitudes of some of the other $X_{aj}$ grow too large, violating the assumptions we made in the beginning on this section.

Fig.~\ref{X4_X5} also reveals that, in the limit where the magnitudes of the $X_{a4}$'s are small, those of the $X_{a5}$'s are ``large.'' This implies that the bounds discussed in the $m_4\gg m_5$ limit above are more stringent if $m_5$ driven oscillations are also accessible to neutrino oscillation experiments. It is, unfortunately, cumbersome and less than illuminating to discuss bounds on the low-energy seesaw with two light right-handed neutrinos. The reasons are simple to understand. For example, one needs to discuss constraints as a function of both $m_4$ and $m_5$. Furthermore, oscillation lengths depend not only on $\Delta m^2_{i4}$ and $\Delta m^2_{i5}$, $i=1,2,3$ but also on $\Delta m^2_{45}$, while the oscillation amplitudes will depend on several different combinations of the $U_{a4,5}$, including relative phases. This all implies that, even though there are upper bounds on the magnitudes and relative phases of the elements of $U_{a4,5}$, they depend non-trivially on $m_4$ and $m_5$. 
 
Here, instead, we will discuss the limiting case where $m_4$ and $m_5$ are very similar. In this case, all short-baseline oscillation probabilities will depend, approximately, on effective mixing angles, given by 
\begin{eqnarray}
&\sin^22\vartheta_{aa}=4\left(|U_{a4}|^2+|U_{a5}|^2\right), \\
&\sin^22\vartheta_{ab}=4\left|U_{a4}U^*_{b4}+U_{a5}U^*_{b5}\right|^2.
 \end{eqnarray}
 where $a,b=e,\mu,\tau$ and $a\neq b$. It is easy to see from Fig.~\ref{X4_X5} that none of the $\vartheta_{aa}$ can vanish, regardless of the mass hierarchy. Numerically, defining $m_4=m_5=M$, expectations for disappearance experiments are
 \begin{eqnarray}
 \sin^22\vartheta_{ee} \ge 0.009\times \frac{\rm 1~eV}{M} & {\rm or} & \ge 0.08\times \frac{\rm 1~eV}{M}, \\
 \sin^22\vartheta_{\mu\mu}  \ge 0.08\times \frac{\rm 1~eV}{M} & {\rm or} & \ge 0.02\times \frac{\rm 1~eV}{M}, \\
 \sin^22\vartheta_{\tau\tau}  \ge 0.08\times \frac{\rm 1~eV}{M} & {\rm or} & \ge 0.02\times \frac{\rm 1~eV}{M}, 
 \end{eqnarray}
 for a normal or inverted neutrino mass hierarchy. Keep in mind that, as in the $m_4\gg m_5$ case discussed above, these bounds cannot all be saturated at the same time. Hence, for $M$ values less than 10~eV, one should observe $\nu_{\mu}$ ($\nu_e$) disappearance at more than the one percent level in the case of a normal (inverted) neutrino mass hierarchy.
   
In the case of appearance experiments, the following upper bound applies:
\begin{eqnarray}
 \sin^22\vartheta_{\mu\tau}  \ge 1.7\times 10^{-3}\times \left(\frac{\rm 1~eV}{M}\right)^2 & {\rm or} & \ge 3\times 10^{-4}\times \left(\frac{\rm 1~eV}{M}\right)^2, 
\end{eqnarray}
for a normal or inverted neutrino mass hierarchy. No nontrivial bound exists for $\sin^22\vartheta_{e\mu}$ and $\sin^22\vartheta_{e\tau}$. All lower bounds are consistent with the current short baseline data for any $M>1$~eV. Expectations for $\nu_{\mu}\to\nu_{\tau}$ appearance would be challenged by next-generation searches for short baseline $\nu_{\tau}$ appearance from conventional neutrino superbeams for $M$ values less than several eVs, for either mass hierarchy. As before, the different inequalities cannot all be saturated at once so the combination of results from different searches is more restrictive than the bounds of each individual search.
  
\subsection{$n=3$}
  
In the case of three right-handed neutrinos, $X$ is a $3\times 3$ matrix. Compared to the $n=2$ case, the number of physical parameters we need to consider is much larger. As far as the ``active'' parameters are concerned, the lightest neutrino mass is no longer zero (but constrained to be less than 0.3~eV or so \cite{Nakamura:2010zzi}) and there are two physical Majorana phases, $\psi$ and $\phi$, regardless of the mass hierarchy. The $R$ matrix (see Eq.~(\ref{eq:casas_ibarra})) is now a complex $3\times 3$ orthogonal matrix parameterized by 3 complex angles (here $\zeta,\eta,\xi$), or six real parameters. The analysis of the allowed values of $X$ is much more cumbersome, and the results one can obtain are much less constrained. For this reason, our discussion will be briefer and less quantitative. Nonetheless, we will offer some qualitative statements that will prove useful to understanding the potential impact of next-generation experiments and how they might test the $n=3$ scenario in the future.

In the case $m_4\ll m_5,m_6$, neutrino oscillation experiments are only sensitive to oscillations involving the $\nu_4$ state. The relevant entries in $X$ can be written as
\begin{eqnarray}
X_{e4}=(V_{\rm MNS})_{e1}\sqrt{\frac{m_1}{m_3}}\cos\zeta\cos\xi + (V_{\rm MNS})_{e2}\sqrt{\frac{m_2}{m_3}}\sin\zeta\cos\xi + (V_{\rm MNS})_{e3}\sin\xi, \\
X_{\mu4}=(V_{\rm MNS})_{\mu1}\sqrt{\frac{m_1}{m_3}}\cos\zeta\cos\xi + (V_{\rm MNS})_{\mu2}\sqrt{\frac{m_2}{m_3}}\sin\zeta\cos\xi + (V_{\rm MNS})_{\mu3}\sin\xi, \\
X_{\tau4}=(V_{\rm MNS})_{\tau1}\sqrt{\frac{m_1}{m_3}}\cos\zeta\cos\xi + (V_{\rm MNS})_{\tau2}\sqrt{\frac{m_2}{m_3}}\sin\zeta\cos\xi + (V_{\rm MNS})_{\tau3}\sin\xi,
\end{eqnarray} 
in the case of a normal neutrino mass hierarchy, and
\begin{eqnarray}
X_{e4}=(V_{\rm MNS})_{e1}\sqrt{\frac{m_1}{m_2}}\cos\zeta\cos\xi + (V_{\rm MNS})_{e2}\sin\zeta\cos\xi + (V_{\rm MNS})_{e3}\sqrt{\frac{m_3}{m_2}}\sin\xi, \\
X_{\mu4}=(V_{\rm MNS})_{\mu1}\sqrt{\frac{m_1}{m_2}}\cos\zeta\cos\xi + (V_{\rm MNS})_{\mu2}\sin\zeta\cos\xi + (V_{\rm MNS})_{\mu3}\sqrt{\frac{m_3}{m_2}}\sin\xi, \\
X_{\tau4}=(V_{\rm MNS})_{\tau1}\sqrt{\frac{m_1}{m_2}}\cos\zeta\cos\xi + (V_{\rm MNS})_{\tau2}\sin\zeta\cos\xi + (V_{\rm MNS})_{\tau3}\sqrt{\frac{m_3}{m_2}}\sin\xi,
\end{eqnarray} 
 in the case of an inverted one. In either case, it is clear that one can pick values for the complex angles $\zeta$ and $\xi$ such that two of the three $X_{a4}$ are vanishingly small. This renders this scenario impossible to definitively rule out in practice for any value of $m_4$ (as long as $V_{\rm MNS}$ is unitary ``enough''). If $\zeta$ and $\xi$ are such that $X_{e4}$ and $X_{\mu4}$ vanish, the only way to constrain this model would be to search for $\nu_{\tau}$ disappearance. Needless to say, experimentally, we are very far away from being able to do that with any precision.
 
 It is also possible to choose parameters such that all $X_{a4}$ are very small. In the case of a normal hierarchy, this can be achieved if $m_1\ll m_2$ and both $\sin\xi$ and $\sin\zeta$ are very small in magnitude. In the case of an inverted hierarchy, the same effect can be obtained if $m_3\ll m_2$ and $|\cos\xi|\ll1$. Under these circumstances, qualitatively, the lightest mostly sterile neutrino ``couples'' predominantly with the lightest mostly active neutrino ($\nu_1$ or $\nu_3$, depending on the mass hierarchy) and effectively decouples from the observable world in the limit where the lightest mass --- currently unconstrained by experiment --- is very small. In the limit that the lightest mass vanishes and this state becomes completely invisible,\footnote{In the case of an inverted mass hierarchy, this would happen when $\cos\xi=0$ and $m_3=0$.} we are back to the $n=2$ case where $\nu_5$ plays the role of $\nu_4$ and $\nu_6$ that of $\nu_5$. Very small values for all $X_{a4}$ are required if one wants the lightest mostly sterile state to play the role of warm dark matter, as discussed in \cite{Asaka:2005an}. 
 
\begin{figure}
\includegraphics[width=0.6\textwidth]{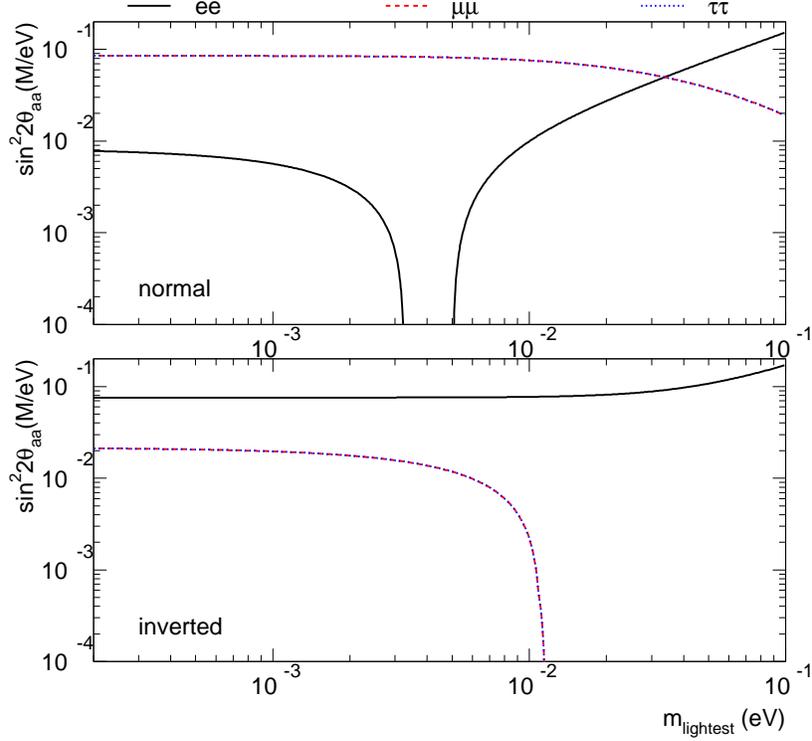}
\caption{Lower bounds for $\sin^22\vartheta_{ee}\times (M/1~\rm eV)$ (black, solid line), $\sin^22\vartheta_{\mu\mu}\times (M/1~\rm eV)$ (red, dashed line), and $\sin^22\vartheta_{\tau\tau}\times (M/1~\rm eV)$ (blue, dotted line) as a function of the smallest neutrino mass $m_{\rm lightest}$ for a normal (TOP) and inverted mass hierarchy (BOTTOM), for $n=3$ and $m_4=m_5=m_6=M$.}
\label{fig:33}
\end{figure}

 In the case $m_4\sim m_5\sim m_6$, the situation is much more constrained. As in the $n=2$ case, we will restrict our attention to effective disappearance and appearance mixing parameters, 
 \begin{eqnarray}
&\sin^22\vartheta_{aa}=4\left(|U_{a4}|^2+|U_{a5}|^2+|U_{a6}|^2\right), \\
&\sin^22\vartheta_{ab}=4\left|U_{a4}U^*_{b4}+U_{a5}U^*_{b5}+U_{a6}U^*_{b6}\right|^2.
 \end{eqnarray}
 We performed a scan of the unconstrained part of the parameters space. In the $n=3$ case, there are ten parameters: three complex mixing angles in $R$ ($\zeta,\eta,\xi$), one Dirac and two Majorana phases in $V_{\rm MNS}$ ($\delta,\psi,\phi$), and the smallest neutrino mass ($m_1$ or $m_3$ for a normal or inverted hierarchy) while keeping the other parameters fixed at their predetermined values, as described in Eq.~(\ref{MNS_data}) and the text that surrounds it. Figure~\ref{fig:33} depicts lower bounds for $\sin^22\vartheta_{aa}\times (M/1~\rm eV)$ as a function of the smallest neutrino mass $m_{\rm lightest}$\footnote{$m_{\rm lightest}\equiv m_1$ in the case of a normal hierarchy, $m_{\rm lightest}\equiv m_3$ in the case of an inverted one.} for both normal and inverted mass hierarchies. Similar upper bounds for $\sin^22\vartheta_{ab}$ prove to be less useful. The reason is that cancellations among the three different contributions to $\sin^22\vartheta_{ab}$ allow for very small lower bounds that are, for all practical purposes, vanishingly small. 

Figure~\ref{fig:33} reveals that, for most values of $m_{\rm lightest}$, the lower bounds here are similar to the ones obtained in the $n=2$ case when $m_4=m_5$. In fact, for $m_{\rm lightest}\lesssim 10^{-3}$~eV, values much smaller than those of the other neutrino masses, the lower bounds are virtually the same. For intermediate values of $m_{\rm lightest}$, ``cancellations'' can occur. These, it turns out, are somewhat familiar to those well-versed in the neutrino physics literature.  

The lower bounds depicted in Figure~\ref{fig:33} are very simple to understand. When $m_4=m_5=m_6=M$, Eq.~(\ref{eq:0const}), for $a=b$, translates into
\begin{equation}
\left(U^2_{a4}+U^2_{a5}+U^2_{a6}\right)M=\sum_{i=1}^3 (V_{\rm MNS})_{ai}(V_{\rm MNS})_{ai} m_i\equiv m_{aa}.
\end{equation}
Keep in mind that the different terms above are, in general, complex. This, in turn, implies 
\begin{equation}
4\left(|U_{a4}|^2+|U_{a5}|^2+|U|^2_{a6}\right)=\sin^22\vartheta_{aa}\ge4\frac{|m_{aa}|}{M}.
\end{equation}
One way to saturate the bound is to choose $\xi,\eta,\zeta$ such that two of the $U_{a4,5,6}$ vanish. Some care is required when $m_{aa}$ vanishes exactly (which happens for a range of values of $m_{\rm lightest}$ for either hierarchy, see Figure~\ref{fig:33}), but it is safe to state that, under this circumstance, the lower bounds for some $\sin^22\vartheta_{aa}$ are vanishingly small for all practical purposes. The same argument also holds in the $n=2$ case, as briefly discussed in the previous subsection. There, none of the $|m_{aa}|$ values are allowed to vanish. 

$m_{aa}$, curiously enough, but not surprisingly, is the magnitude of the $aa$-element of what can be refereed to as the ``active--active neutrino mass matrix,'' $m_{ab}$. $m_{ab}$ is the effective neutrino mass matrix in the limit of very heavy sterile neutrino masses, and is also the active Majorana neutrino mass matrix that is generated in the so-called type-II seesaw (in a nutshell, the Standard Model Lagrangian augmented by an SU(2) triplet scalar that acquires a vacuum expectation value). The attentive reader will have noticed that the lower bound curve for $\sin^22\theta_{ee}$ is exactly proportional to the lower bounds on the magnitude of $m_{ee}$ (often referred to as $m_{\beta\beta}$ in the literature), used to estimate the sensitivity of neutrinoless double-beta decay experiments (see, for example, \cite{Nakamura:2010zzi}). The equivalent observable with muon neutrinos, the magnitude of $m_{\mu\mu}$, is less familiar but has been studied in the literature along with the magnitude of the other elements of $m_{ab}$ \cite{Merle:2006du}. In the Appendix, we plot the allowed values of the magnitude of $m_{\mu\mu}$, for both mass hierarchies, as a function of $m_{\rm lightest}$, for the values of the mixing parameters chosen here. It is curious that $m_{\mu\mu}$ vanishes only for an inverted mass hierarchy, and only in the limit when $m_3\sim m_2$ (to be more precise, $m_3\gtrsim m_2/3$). This is very easy to understand from the definition of $m_{\mu\mu}$ and the known values of the active neutrino parameters \cite{Merle:2006du}. 
  
Finally, in the $n=3$ case, one may also consider the case $m_4\lesssim m_5 \ll m_6$. Qualitatively, we anticipate that the situation will resemble the $n=2$ case, especially in the limit where we treat $\nu_4$ and $\nu_5$ as one ``effective'' state. Scenarios of this type may be phenomenologically interesting. For example, $\nu_4$ and $\nu_5$ may provide an explanation to the reactor anomaly, while $\nu_6$ may play the role of warm dark matter, as discussed, for example, in \cite{deGouvea:2006gz}.

\setcounter{equation}{0} \setcounter{footnote}{0}
\section{Concluding Statements and Summary}
\label{sec:conclusions}

We have explored the seesaw Lagrangian (Eq.~(\ref{eq:lnu}) in the limit $M_i\gg y^{\alpha i}v$), concentrating on the values of the active--sterile neutrino mixing parameters. The key point is that even though the neutrino mixing matrix $U$ is $(3+n)\times(3+n)$ and unitary, it is not a generic unitary matrix. Several of its parameters are not observable and can be ``rotated away,'' while the remaining parameters are related to one another and to the neutrino mass eigenvalues. We take advantage of these relations in order to ask whether the current and the next generation of short-baseline neutrino experiments is capable of definitively ruling out --- or, perhaps, ruling ``in'' --- eV-scale right-handed neutrino masses for all allowed values of the unknown parameters in $U$. 

If there are only two right-handed neutrinos ($n=2$, fittingly dubbed the minimal model in \cite{Donini:2011jh}), the situation is quite constrained, especially if the neutrino mass hierarchy is inverted. In this case, short baseline neutrino experiments should be able to  rule out $m_4$ values less than several eV regardless of the value of $m_5$ as long as one is sensitive to several different oscillation channels. This statement is especially true if one can probe $\nu_e$ and $\nu_{\mu}$ disappearance at the few percent level, and $\nu_e$ and $\nu_{\tau}$ appearance at the $10^{-4}$ level or better. For $m_4$ values above 10~eV, the low energy seesaw is harder to unambiguously test. A potentially interesting channel to pursue is $\nu_{\tau}$ appearance at the $10^{-5}$ level \cite{Alonso:2010wu} or better. 

On the flip side, if light sterile neutrinos do manifest themselves in next generation experiments, it is, in principle, possible to determine whether these sterile neutrinos are described by Eq.~(\ref{eq:lnu}). Again, in order to do that, one must probe several different oscillation channels. In the concrete scenarios spelled out in Eqs.~(\ref{eq:fit_normal},\ref{eq:fit_inv}), not only did we reproduce (close to) the best fit values for the 3+2 fit to current short-baseline data (see \cite{Kopp:2011qd}), we also ``predict'' that $|U_{\tau 4,5}|$ is at least as large as $|U_{\mu4,5}|$ and $|U_{e4,5}|$. This, in turn, would imply that $\nu_{\mu,e}\leftrightarrow\nu_{\tau}$ oscillations must be observed at a level similar to, or larger than,  $\nu_{\mu}\leftrightarrow\nu_{e}$ oscillations.

For three or more right-handed neutrinos, life is harder. Even if, say, $m_4=1$~eV, $m_5=10$~eV, and $m_6=100$~eV, it is possible to ``hide'' the $\nu_4$ state from the active neutrinos at a level such that oscillations mediated by $m_4^2$ are severely suppressed. Null results can still constrain the parameter space, but the constraints are rather weak even for very light right-handed neutrinos, as long as their masses are hierarchical.   

We restricted our discussion to neutrino oscillation experiments. The main reason is that these have the most sensitivity to eV-scale sterile neutrinos and, as already summarized above, oscillations allow one to probe different combinations of elements of the mixing matrix. For eV sterile neutrinos, however, precision measurements of $\beta$-decay spectra and searches for neutrinoless double beta decay ($0\nu\beta\beta$) can also provide powerful information \cite{deGouvea:2005er,deGouvea:2006gz,Barrett:2011jg,Blennow:2010th}. In both cases, one is sensitive to the $U_{ei}$ elements and the neutrino masses. A detailed estimate of the sensitivity of Katrin (tritium $\beta$-decay) can be found in \cite{Barrett:2011jg}. In the case of $0\nu\beta\beta$, life is more interesting.  It has been pointed out that if all right-handed neutrinos are light (masses below a few MeV), Eq.~(\ref{eq:lnu}) predicts that the rate for $0\nu\beta\beta$ is vanishingly small \cite{deGouvea:2005er,deGouvea:2006gz} (see also \cite{Blennow:2010th,Mitra:2011qr}). A positive signal for $0\nu\beta\beta$ would definitively exclude the possibility that Eq.~(\ref{eq:lnu}) is correct {\sl and} that all right-handed neutrino masses are light.  On the other hand, a negative signal for $0\nu\beta\beta$ combined with the discovery of light sterile neutrinos in neutrino oscillation experiments, and the discovery that the neutrino mass hierarchy is inverted, would provide a very strong boost for the low-energy seesaw scenario.\footnote{Alas, the possibility that the neutrinos are Dirac fermions would, of course, remain.}

Finally, the existence of eV scale sterile neutrinos is challenged by cosmology, especially big bang nucleosynthesis and the large scale structure of the universe (for recent detailed analyses, see \cite{Hamann:2010bk,Hamann:2011ge}).  We have nothing to add to this particular discussion. None too pleasing ``ways out'' have been proposed in in the literature (for many more details, see, for example, \cite{Cirelli:2004cz} and references therein). Instead, we choose to view this issue through an optimistic lens: if eV scale sterile neutrinos are observed in laboratory experiments, our current understanding of the evolution of the universe will also have to change significantly. 

For larger right-handed neutrino masses, the lower bounds on $X_{a4,5,6}$ discussed here translate into very small upper bounds on $U_{a4,5,6}$. Cosmology and astrophysics still imply interesting, nontrivial constraints on the low-energy seesaw models (see, for example, \cite{deGouvea:2006gz,Kusenko:2009up}) up until masses close to 100~keV, and there remains the possibility that one of the right-handed neutrinos is the dark matter, as proposed in \cite{Asaka:2005an}, as long as all $X_{aj}$ are small enough for the ``dark matter'' neutrino $\nu_j$. Terrestrial experiments, however, lack the sensitivity to unambiguously test the low-energy seesaw in case all right-handed neutrino masses are much larger than 1~eV and the lower bounds presented here are saturated.  Indeed, in order to observe the seesaw neutrinos for such ``heavy'' right-handed neutrino Majorana mass parameters, one needs Nature to pick points in the parameter space such that $X_{a4,5,6}$ are much larger than 1 and hence very far from the lower bounds discussed here. This possibility has been considered in the literature (see, for example, \cite{Kersten:2007vk,de Gouvea:2007uz,Ibarra:2011xn}). 

There are many other consequences of the non-generic properties of $U$. Before concluding, we highlight the apparent double-life of the Majorana phases.\footnote{We thank Serguey Petcov for pointing this out at several occasions in the context of leptogenesis and expectations of the minimal supersymmetric standard model for charged-lepton flavor violation. See, for example, \cite{Petcov:2006pc}.} In the case of two active and two sterile neutrinos, Eq.~(\ref{eq:2+2}) expresses the values of $U_{a,b;4,5}$ in terms of the neutrino mass eigenvalues, the active mixing angle $\vartheta$, the complex angle $\zeta$, and the active Majorana phase $\varphi$. Measurements of all oscillation channels would reveal, for a generic choice of the parameters, that $P_{ab}\neq P_{ba}$ and the neutrino and antineutrino appearance probabilities would be different, i.e., that CP-invariance and that T-invariance are both violated. In turn, we would be able to measure the CP-violating parameters $\varphi$ and $\zeta$. This would imply that by exploring only lepton number conserving processes one would be able to determine the Majorana phase or, equivalently, the Majorana phase is responsible for a lepton-number conserving phenomenon,\footnote{There is no theorem that prevents this from happening. For a simple concrete example of Majorana phases impacting CP-violating but lepton-number conserving phenomena, see \cite{deGouvea:2005jj}.} i.e., it determines the magnitude of what is normally referred to as a Dirac phase. While this appears puzzling, it is only a reflection of the way we chose to parameterize the mixing matrix and the neutrino masses. Clearly, the definition of ``input'' and ``output'' parameters is a matter of taste, not physics. Another naive, but much more serious consequence one would be tempted to draw from such a measurement is that the neutrinos are Majorana fermions. The logic is simple. If the neutrinos were Dirac fermions, the Majorana phase would be unphysical and hence the measurement of a Majorana is only possible if the neutrinos are their own antiparticles. The logic is simple, but flawed. One can only infer that neutrinos are Majorana fermions because one has made the assumption that the $2+2$ scenario is correct. There is another scenario, with two Dirac active neutrinos and two Dirac sterile neutrinos, that exactly mimics the $2+2$ scenario above as far as all lepton-number conserving observables are concerned. It is true that the $2+2$ seesaw would be a much more palatable fit to the data --- it would, for example, explain several relations that would appear to be accidental in the case of Dirac neutrinos --- but the evidence for the Majorana nature of the neutrino would remain circumstantial until lepton number violating phenomena were observed. 

We conclude by re-emphasizing the main point of our analysis: the type-I seesaw Lagrangian yields a non-generic set of active--sterile oscillation parameters --- the masses and mixing angles are, in some sense, entwined. For this reason one is capable to, in principle, test the model by performing enough measurements of neutrino oscillations. 

\section*{Acknowledgments}

AdG thanks Serguey Petcov for inspiring conversations, and we thank Alejandro Ibarra for comments and questions regarding the manuscript. This work is sponsored in part by the DOE grant \# DE-FG02-91ER40684.

\appendix

\section{Allowed values for the $\mu\mu$ element of the active neutrino mass matrix}

\begin{figure}
\includegraphics[width=0.6\textwidth]{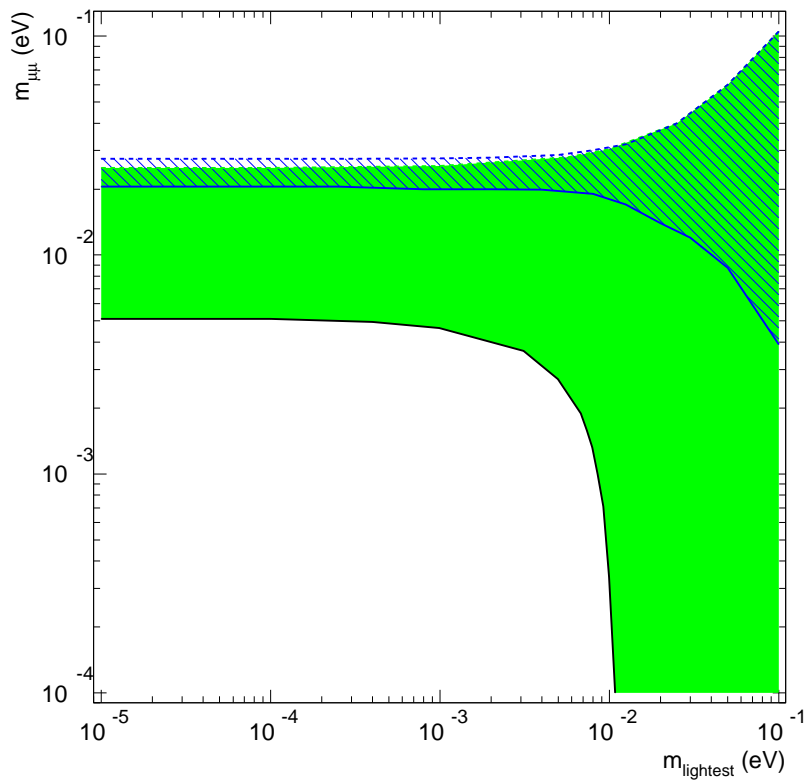}
\caption{Allowed values of the magnitude of $m_{\mu\mu}$ as a function of the smallest neutrino mass $m_{\rm lightest}$, assuming the known oscillation parameters are fixed to Eq.~(\ref{MNS_data}) and allowing all phases to vary within their allowed parameter space, for a normal (blue, open hatched region) or inverted (green, solid region) neutrino mass hierarchy. The solid lower bounds are proportional to the curves presented in Fig.~\ref{fig:33}.}
\label{mmm}
\end{figure}

If the right-handed neutrinos are infinitely heavy and integrated out of the theory, one is left with a $3\times 3$ ``active'' neutrino mass matrix. Its elements are given by
\begin{equation}
m_{ab}=\sum_{i=1,2,3} (V_{MNS})_{ai}(V_{\rm MNS})_{bi}m_i.
\end{equation}
Under these circumstances, the $ee$ element of $m_{ab}$, $m_{ee}$, is the physical quantity probed by $0\nu\beta\beta$ searches and the subject of a lot of theoretical scrutiny (see, for example, \cite{Nakamura:2010zzi}). The $\mu\mu$ element, $m_{\mu\mu}$, plays a similar role for lepton-number violating phenomena involving muon-neutrinos, for example, the very rare lepton-number violating kaon decay $K^{\pm}\to\pi^{\mp}\mu^{\pm}\mu^{\pm}$. For a more detailed discussion see, for example, \cite{Frigerio:2002fb,Merle:2006du}. Figure~\ref{mmm} depicts the allowed values of the magnitude of $m_{\mu\mu}$ as a function of the smallest neutrino mass $m_{\rm lightest}$, assuming the known oscillation parameters are fixed to Eq.~(\ref{MNS_data}) and allowing all phases to vary within their allowed parameter space, for both a normal and inverted neutrino mass hierarchy. Similar plots for all $m_{ab}$ can be seen in \cite{Merle:2006du}.

 \end{document}